\documentstyle[preprint,%amsfonts,%
eqsecnum,aps,prl]{revtex}
\def\scri{{\cal I}}

\def\endproof{\nobreak\kern5pt\nobreak\vrule height4pt width4pt depth0pt}

\begin{document}
\draft

\hyphenation{
mani-fold
mani-folds
geo-metry
geo-met-ric
}

\title{The AdS/CFT Correspondence Conjecture and Topological Censorship}

\author{G. J. Galloway\footnote{e-mail: galloway@math.miami.edu}}
\address{Dept. of Mathematics and Computer Science\\ University of Miami\\
Coral Gables, FL 33124, USA}
\author{K. Schleich\footnote{e-mail: schleich@noether.physics.ubc.ca}}
\address{Department of Physics and Astronomy\\ University of British Columbia\\
Vancouver, BC V6T 1Z1, Canada}
\author {D. M. Witt\footnote{e-mail: donwitt@noether.physics.ubc.ca}}
\address{Department of Physics and Astronomy\\ University of British Columbia\\
Vancouver, BC V6T 1Z1, Canada}
\author{E. Woolgar\footnote{e-mail: ewoolgar@math.ualberta.ca}}
\address{Dept. of Mathematical Sciences and Theoretical Physics Institute\\
University of Alberta\\ Edmonton, AB T6G 2G1, Canada}

\maketitle
\vfill\eject
\begin{abstract}

In \cite{sw:gallowayetal99} it was shown that
$(n+1)$-dimensional
asymptotically anti-de-Sitter spacetimes obeying natural causality
conditions
exhibit topological censorship.  We use this fact in this paper to derive in arbitrary
dimension
relations between the topology of the timelike boundary-at-infinity, $\scri$, and that
of the spacetime interior to this boundary.
We  prove as a simple corollary of topological censorship that any
asymptotically
anti-de Sitter spacetime  with a disconnected boundary-at-infinity
necessarily contains black hole horizons which screen the boundary components
from each other. This corollary  may be viewed as
 a Lorentzian analog of the  Witten and Yau result
 \cite{WittenYau99}, but is independent of the scalar curvature of $\scri$.
Furthermore,  as shown in \cite{sw:gallowayetal99}, the topology of $V'$, the Cauchy surface (as
defined for asymptotically anti-de Sitter spacetime with boundary-at-infinity) 
for regions
exterior to event horizons, 
is constrained by that
of $\scri$;   
 the homomorphism $\Pi_1(\Sigma_0)\to \Pi_1(V')$ induced by 
the inclusion map is onto where $\Sigma_0$ is the intersection of $V'$ with 
$\scri$.
  In $3+1$ dimensions, the homology of $V'$ can be completely determined from this as shown 
in \cite{sw:gallowayetal99}. In this paper, we prove in arbitrary dimension that 
 $  H_{n-1}( V;Z)=Z^k$ where $V$ is the closure of $V'$ and
 $k$ is the number of boundaries $\Sigma_i$ interior to $\Sigma_0$. As a consequence, $V$
does not contain any wormholes or other compact, non-simply connected topological structures.
Finally, for the case of 
$n=2$, we show that these constraints and the onto homomorphism of the fundamental groups
from which they follow
 are sufficient to limit the topology of 
interior of $V$ to either $B^2$ or $I\times S^1$. 
\end{abstract}


\narrowtext

\section{Introduction}
The global structure and topology of asymptotically anti-de Sitter spacetime is
a topic  of particular
 interest due to its relevance to
string theory. Spacetimes that are products of an asymptotically anti-de Sitter spacetime in
$n+1$ dimensions and
a Riemannian manifold arise in the  low
energy limit of certain D-brane configurations, notably for the cases of $n=2$ and 
$n=4$ \cite{dbranerefs}. Furthermore, Maldacena has proposed that supergravity in an 
asymptotically anti-de Sitter
spacetime corresponds to a conformal field theory on  the 
boundary-at-infinity\footnote{Concrete discussions of the causal structure of a 
spacetime begin by relating
a spacetime with infinitely far away
points to one in which all points are at finite distance. This relation is 
carried out by finding both a coordinate transformation
and a conformal factor such that the original spacetime metric is conformally 
related to a metric with finite coordinate ranges on a spacetime-with-boundary 
(cf. \cite{HawkingandEllis}).
The boundary-at-infinity refers to the boundary of this conformally related spacetime.}
of this spacetime
in the large $N$ limit \cite{maldacena}. This conjecture, the adS/CFT correspondence conjecture, 
is  supported by recent calculations which, for example, show a direct
connection between
black hole entropy as calculated classically and the number of states of the 
conformal field theory on the boundary-at-infinity \cite{recentreview}. Thus, the 
adS/CFT correspondence conjecture
provides new insight into the old puzzle of black hole entropy in the context of
string theory. Moreover, it is believed that this conjecture, if true, may hold answers to
other long-standing puzzles in gravity.

Now it is well known that asymptotically anti-de Sitter spacetimes admit black holes and wormholes of various topologies 
\cite{sw:banados1992,sw:lemos95,sw:aminneborgetal96,mann97,brill97} (for a recent review see
\cite{mannreview}).
These spaces  exhibit a boundary-at-infinity 
 which carries the topology of the event horizons. Furthermore 
one can show that there exist initial data sets with very general topology that evolve as 
 anti-de Sitter spacetimes
 \cite{mess}. 
 
It is therefore natural to ask what implications the topology of an asymptotically
anti-de Sitter spacetime has for the adS/CFT correspondence. That is, if the topology of an 
asymptotically
 anti-de Sitter spacetime is
 arbitrary, can a conformal field theory that only detects the topology of its boundary-at-infinity
  correctly
  describe its physics? Recently,
Witten and Yau in \cite{WittenYau99} have addressed part of this issue in the context of  
a generalization of the
adS/CFT correspondence in which this conjecture is formulated in terms of 
Riemannian manifolds  \cite{witten98}. They show that
  the topology 
of a complete Einstein manifold $M$ of negative curvature and
 boundary $N$ admitting positive scalar curvature is constrained. In particular, 
 they show $H_n(M;Z)=0$ and thus
  $N$ is connected. A consequence of this result is that
  Riemannian manifolds satisfying these conditions
 do not admit wormholes.

However, the interesting results in \cite{WittenYau99} do not address the relation of the 
topology   
of an asymptotically anti-de Sitter space
to that of the boundary-at-infinity in the context of Lorentzian spacetimes,
the standard arena for the adS/CFT correspondence
conjecture. We will address this issue. We will do so by exploiting the fact that asymptotically
anti-de Sitter spacetimes are members of a class of spacetimes that exhibit
{\it topological censorship}; that is
any causal curve with initial and final endpoints on the boundary-at-infinity $\scri$ 
can be continuously deformed
to a curve that lies in $\scri$ itself. Thus causal curves passing through the interior
of an asymptotically anti-de Sitter spacetime detect no topological structure not also present in the boundary-at-infinity.
This somewhat surprising result was first proven to hold for  asymptotically 
flat spacetimes by Friedman, Schleich
and Witt\cite{sw:friedmanetal93}. It has been generalized to apply to a broader class of spacetimes 
\cite{sw:gallowayetal99,sw:galloway96,sw:gallowaywoolgar97}. 
Of relevance here is the proof in \cite{sw:gallowayetal99} 
that topological censorship holds for a class of spacetimes with
timelike boundary-at-infinity that includes asymptotically anti-de Sitter spacetimes. This
proof, like other proofs of topological censorship,
holds in any such $(n+1)$-dimensional spacetime with  $n\ge 2$. 
This fact
 immediately implies that causal curves passing 
through the interior of an asymptotically anti-de Sitter spacetime detect only topological structure also 
present in the boundary-at-infinity in any such dimension.

It is also known that more
 information about the topology of a spacetime  that exhibits topological censorship can be found by
algebraic topology. In $3+1$ dimensions,
Chru\'sciel and Wald \cite{ChruWald} noted that topological censorship
implies that black holes are topological 2-spheres in  stationary,
asymptotically flat spacetimes.
Jacobson and Venkataramani 
\cite{sw:jacobsonetal95} generalized this result to show that it holds 
for a quite general class of $(3+1)$-dimensional asymptotically flat spacetimes, including
spacetimes with black hole formation by collapse.
For the case of $(3+1)$-dimensional asymptotically anti-de Sitter spacetimes, 
we showed in \cite{sw:gallowayetal99} that the sum of the genera of black 
holes is bounded above by that of a spatial cut of
the boundary-at-infinity.  Furthermore, the
 integral homology of 
 Cauchy surfaces (as
defined for asymptotically anti-de Sitter spacetime with boundary-at-infinity) is torsion free
and consequently completely determined by the Betti numbers. Moreover,
the first Betti number is equal to the sum of the genera of the
black holes and the second Betti number is equal to the number of black holes.
Thus topological censorship
 restricts both the topology of black holes and
that of the spacetime exterior to them in $3+1$ dimensions.

Though the proof of topological censorship in \cite{sw:gallowayetal99} holds in $(n+1)$-dimensional
spacetime for 
 $n\ge 2$,
  some of the
stronger results in \cite{sw:gallowayetal99} have been derived  by using
certain special properties of
the topology of  $(3+1)$-dimensional spacetimes. Clearly
  one can  extend  some of these arguments using algebraic topology 
to  other dimensions. 
 We will do so in this paper.
 
 We first prove a simple corollary of topological censorship, that any asymptotically 
anti-de Sitter spacetime  with a disconnected boundary-at-infinity 
necessarily 
contains black hole horizons which screen the boundary components
from each other. This result is independent of the scalar curvature of the
boundary-at-infinity. But, in a certain sense, it is a Lorentzian analog of 
the  Witten and Yau result
 \cite{WittenYau99}.
Furthermore the topology of $V'$, the  Cauchy surface (as
defined for asymptotically anti-de Sitter spacetime with boundary-at-infinity) for regions
exterior to event horizons 
is constrained by that
of the boundary-at-infinity; we show that
 the homomorphism $\Pi_1(\Sigma_0)\to \Pi_1(V')$ induced by 
the inclusion map is onto where $\Sigma_0$ is the intersection of $V'$ with the  
boundary-at-infinity. We also prove that the  integral homology  $H_{n-1}( V;Z)=Z^k$ where
$ V$ is the closure of $V'$
and $k$ is the number of boundaries $\Sigma_i$ interior to $\Sigma_0$. As a consequence $V$ itself
does not contain any wormholes or other compact non-simply topological structures.
Furthermore, the integral homology $H_k(V;Z)$ is 
torsion free for $k= n-2$.  For the case of 
$n=2$, these constraints and the onto homomorphism of the fundamental groups
are sufficient to limit the topology of 
interior of $V$ to either $B^2$ or $I\times S^1$. Therefore, in $2+1$ dimensions, the
topology of the boundary-at-infinity almost completely characterizes that of the interior, a
desirable conclusion for the adS/CFT correspondence conjecture. However, in $4+1$ dimensions, 
these
constraints and the onto homomorphism of fundamental groups
 in and of themselves are not sufficient to completely characterize the topology of the
interior. As we will see, they do not suffice to constrain the number of compact
simply connected topological structures in the interior.

\section{Topological censorship in $(n+1)$-dimensional asymptotically anti-de Sitter spacetimes} 

Precisely, we will consider an $(n+1)$-dimensional connected spacetime 
$\cal M$, with metric $g_{ab}$, which can be
conformally included into a  spacetime-with-boundary
$\cal M'= \cal M \cup \scri$, with metric $g_{ab}'$, such that $\partial 
\cal M' = \scri$ is timelike ({\it i.e.}, is an $n$-dimensional Lorentzian hypersurface 
in the induced metric) and $\cal M = \cal M' \setminus \scri$. Note
that the boundary-at-infinity $\scri$ can have multiple components, 
that is the cardinality of $\Pi_0(\scri)$ can
be greater than one. 
The conditions on the
conformal factor
 $\Omega\in C^{1}(\cal M')$ are that
(a) {$\Omega > 0$ and $g_{ab}' = \Omega^2 g_{ab}$  on $\cal M$, and}
(b) {$\Omega = 0$ and $d \Omega \ne 0$ pointwise on $\scri$.}
These are the standard conditions on $\Omega$ in a conformal compactification
of a spacetime with infinitely extendible null geodesics such as asymptotically 
anti-de Sitter spacetimes. 

The conformal compactification of  the universal anti-de Sitter spacetime in $n+1$ dimensions
(cf. \cite{HawkingandEllis} p. 131) is a canonical example of such a spacetime.   
Its boundary-at-infinity $\scri$ is $n$-dimensional
Minkowski spacetime.\footnote{Precisely, the boundary-at-infinity is the
Einstein static universe $S^{n-1}\times R$ in which $n$-dimensional Minkowski spacetime is
itself conformally included.}
 Group actions on this spacetime  generate local adS spacetimes containing black holes 
and wormholes (see for example
\cite{sw:lemos95,sw:aminneborgetal96}). In contrast to universal anti-de Sitter spacetime, 
these spacetimes can have disconnected $\scri$. 

A spacetime-with-boundary $\cal M'$ is defined to be {\it globally hyperbolic} if
$\cal M'$ is strongly causal and the intersection of the causal future of $p$ with
 the causal past of $q$, $J^+(p,{\cal M'})\cap J^-(q,\cal M')$, is compact for 
 all $p,q\in \cal M'$.\footnote{Note that the causal future
of a set $S$ relative to  $U$, $J^+(S,U)$, is the union of $S\cap U$ with the
set of all points that can be reached from $S$ by a future directed  
non-spacelike curve in $U$. The interchange of the past with future in the previous
definition yields $J^-(S,U)$. }  This definition has exactly the same form as that for the
case of a spacetime without boundary. However, observe that the inclusion of the boundary
is key for spacetimes with timelike $\scri$. In particular,
${\cal M}$,  is not globally hyperbolic  but  ${\cal M}'={\cal M}\cup \scri$ is globally hyperbolic
for many spacetimes with timelike $\scri$ such as asymptotically anti-de Sitter spacetimes.  

For example, it is well known that universal anti-de Sitter spacetime ${\cal M}$
itself is not globally hyperbolic (cf. \cite{HawkingandEllis} p. 132); 
as $\scri$ is timelike,  one can find points $q$ and
$p$ such that $J^+(p,{\cal M})\cap J^-(q,\cal M)$ is not compact. Such points are those
such that past directed radially outward null curves from $q$ and future directed radially
outward null curves from $p$
intersect $\scri$ before they intersect each other. As these geodesics leave
$\cal M$ before intersecting,  $J^+(p,{\cal M})\cap J^-(q,\cal M)$ is not compact.
     However, observe that 
$J^+(p,{\cal M'})\cap J^-(q,\cal M')$ is compact as it
includes the appropriate part of 
 $\scri$. Intuitively, one is including additional information about the
spacetime by including the boundary-at-infinity and this additional information is sufficient
to ensure physical predictability.

We will also require that the spacetime satisfy
 a modified form of the Averaged Null Energy Condition (ANEC):
For each point $p$ in $\cal M$ near ${\scri}$ and any
future complete null
geodesic $s\to\eta(s)$ in $\cal M$ starting at $p$ with tangent $X$, $\int_0^{\infty}{\rm
Ric}(X,X)\,ds\ge
0$.\footnote{ The term ANEC
usually refers to a condition of this form except that the 
integral is taken over geodesics complete to both past and future \cite{borde}.}
This condition is necessary to ensure that all radially outward directed null geodesics 
from a closed outer trapped surface focus
to a conjugate point in finite affine parameter, ensuring that
  this surface is not visible to $\scri$ (see for example
  Prop. 9.2.1 in \cite{HawkingandEllis}, a related proof using the null energy condition).

ANEC is stated in geometric form but can be interpreted physically
by invoking the Einstein equations to relate the Ricci tensor to its sources. 
In particular, if the Einstein equations
with cosmological constant hold,
$R_{ab} - {1\over 2} R g_{ab} + \Lambda g_{ab} = 8\pi T_{ab}$,
 then as $ g_{ab}X^aX^b=0$ for any null vector $X$, 
 ${\rm Ric}(X,X) = R_{ab} X^aX^b 
= 8\pi T_{ab}X^aX^b= 8\pi T(X,X)$. 
Clearly, the 
cosmological constant does not appear in this expression. Consequently, 
ANEC depends only on the stress energy tensor.  ANEC
 is satisfied by the stress energy tensor of physically reasonable sources of matter.
In particular, it is obvious that spacetimes with negative
cosmological constant containing no 
matter will satisfy this condition. 

Finally, a spacetime satisfies the {\it generic condition} if every timelike or null geodesic
with tangent vector $X$
contains a point at which $X^aX^bX_{[c}R_{d]ab[e}X_{f]}$ is not zero.
The generic condition will be satisfied if a spacetime contains matter or gravitational radiation
in a non-symmetric configuration.\footnote{All known examples of spacetimes that do
 not satisfy the generic condition have a high degree of local or global symmetry.}

We begin by reminding readers
that proofs of topological
censorship, in particular that of
 \cite{sw:gallowayetal99}, hold in $(n+1)$-dimensional
spacetimes for $n\ge 2$. Namely,

\noindent {\bf Theorem 1}. {\sl Let ${\cal M}'$ be a globally hyperbolic 
spacetime-with-boundary with timelike boundary $\scri$ that satisfies ANEC. Let 
$\scri_0$ be a connected component of 
$\scri$. Furthermore
 assume either
({\it i}) { ${\scri_0}$ admits a compact spacelike cut}
or 
 ({\it ii}) {$\cal M'$ satisfies the generic condition.}
Then every causal curve 
whose initial and final endpoints belong to ${\scri_0}$ is fixed end point
homotopic to a curve on ${\scri_0}$ }
\vskip 1pt

\noindent
This is an alternate but completely equivalent statement of theorem 2.2 proven 
in \cite{sw:gallowayetal99}.
The proof of theorem 1 uses the result

\noindent {\bf Theorem 2}. {\sl Let ${\cal M}'$ be a globally hyperbolic spacetime-with-boundary
with 
timelike boundary $\scri$ that satisfies ANEC. Let $\scri_0$ be a connected component of 
$\scri$ of  $\cal M'$. Furthermore
 assume that either
({\it i}) { ${\scri_0}$ admits a compact spacelike cut}
or 
 ({\it ii}) {$\cal M'$ satisfies the generic condition.}
Then $\scri_0$ cannot communicate with any other component of $\scri$,
{\it i.e.}, $J^+(\scri_0)\cap
(\scri\setminus \scri_0) = \emptyset$. }
\vskip 1pt
This theorem is a restated form of theorem 2.1  proven 
in \cite{sw:gallowayetal99}. The maximally extended Schwarzchild and Schwarzchild-anti-de Sitter
 solutions
provide simple examples of spacetimes satisfying the conditions of this theorem. They 
both have two disconnected components of
$\scri$. Causal curves originating from one component of $\scri$ 
cannot end on the other; instead
they end on the black hole singularity.

 It is useful to mention that
theorem 1  follows from theorem 2 by  constructing a covering 
space of ${\cal M}'$ in which all
non-contractible curves not homotopic to curves on  $\scri_0$  are unwound.
Any causal curve with endpoints on $\scri_0$
not fixed endpoint homotopic to a causal curve in  ${\scri}_0$ 
will
begin  on a different component of $\scri$ in this covering space.
However, this covering space is itself a globally hyperbolic spacetime-with-boundary
 satisfying the
conditions of theorem 2. Thus such a curve cannot
exist. Hence the result.

Ref. \cite{sw:gallowayetal99}  provides a
 natural restatement of theorem 1 in terms of the region of spacetime that
can communicate with a given component of the boundary-at-infinity. This region, 
the domain of outer communications
${\cal D} = I^-(\scri_0)\cap I^+(\scri_0)$,\footnote{Note that the timelike future
of a set $S$ relative to  $U$, $I^+(S,U)$, is the 
set of all points that can be reached from $S$ by a future directed  
timelike curve in $U$. The interchange of the past with future in the previous
definition yields $I^-(S,U)$. } is the subset of ${\cal M}$ that
is in causal contact with $\scri_0$. 
As we shall see later, one can think of the domain of outer communications
as the region of ${\cal M}$ which is exterior to event horizons. Clearly,
 ${\cal D} $ is also the interior of an $(n+1)$-dimensional spacetime-with-boundary
 ${\cal D}' = {\cal D }\cup \scri_0$. 
Now theorem 1 can be conveniently restated in
terms of the fundamental group of the domain of outer communications. 
Observe 
 that the inclusion map $i:\scri_0 \to \cal D'$ induces a homomorphism
of fundamental groups  $i_*:\Pi_1({\scri_0})\to\Pi_1(\cal D')$. Then
\vskip 1pt
\noindent
{\bf Theorem 3}. {\sl If ${\cal M'}$ is a globally hyperbolic spacetime-with-boundary that
satisfies the conditions given in theorem 1, 
then the group homomorphism
$i_*:\Pi_1({\scri_0})\to\Pi_1({\cal D'})$ induced by inclusion is surjective.}
\vskip 1pt

\noindent
Theorem 3 says roughly that every loop in $\cal D$ is deformable to
a loop in $\scri$.
Moreover, it implies that $\Pi_1({\cal D})$ ($= \Pi_1({\cal D}')$) is isomorphic
to  the factor group $\Pi_1({\scri})/{\rm ker}\,i_*$.  In particular, if
$\scri$ is simply connected
then so is $\cal D$, thus generalizing the result of \cite{sw:galloway95}.

\section{Causal disconnectedness of disjoint components of the boundary-at-infinity} 

The boundary of the region of spacetime visible to
observers at $\scri$ by past directed causal curves 
is referred to as the
future event horizon.
This horizon is a set of one or more null surfaces, also called black hole horizons,
 generated by null geodesics
that have 
no future endpoints but possibly 
have past endpoints. Precisely these horizons are characterized as the boundary of the causal past of
$\scri$, $\dot J^-(\scri)$. A past event horizon is similarly defined; $\dot J^+(\scri)$. The past
event horizon also can consist of one or null surfaces known as white hole horizons.

Theorems 1, 2 and 3 provide a partial characterization of the topology of the
region of spacetime exterior to  the event horizons
in $(n+1)$-dimensions. In particular, they demonstrate that no causal curve
links with  event horizons in a manner such that it cannot be deformed
to a curve on the boundary-at-infinity.
Rather causal curves in the spacetime will only carry information
about the non-triviality of curves on $\scri$. Thus the topology of 
event  horizons in  spacetimes that exhibit topological censorship is constrained.  

An immediate result of theorem 2  is that spacetimes 
with disconnected $\scri$ contain black hole horizons which screen the boundary components
from each other.

\noindent {\bf Corollary}. {\sl Let ${\cal M}'$ satisfy the conditions given in theorem 1. 
If $\scri$ is disconnected, then
the spacetime contains black hole horizons, namely $\dot J^-(\scri_0)\neq \emptyset$.}
\vskip 1pt

{\noindent \bf Proof}: Let  $\scri_1$ be  any component of $\scri$ not connected to $\scri_0$.
Theorem 2 shows that there is no causal curve connecting $\scri_0$ and $\scri_1$.
  Thus the causal past of
$\scri_0$ is disjoint from the causal future of $\scri_1$, $J^-(\scri_0)\cap J^+(\scri_1)=\emptyset$.
 Now as  both are subsets of ${\cal M'}$, clearly
  $J^-(\scri_0)$ is not itself ${\cal M'}$. Thus  $\dot J^-(\scri_0)\neq \emptyset$.\hfill\endproof
  
Observe that a similar argument shows that the spacetime contains a past event horizon,
$\dot J^+(\scri)$;  as $J^+(\scri_1)$ is also not itself ${\cal M'}$,
 $\dot J^+(\scri_1)\neq \emptyset$. Note that $\dot J^-(\scri_0)$ and $\dot J^+(\scri_1)$
 may coincide as is the case in the maximally extended Schwarzchild and Schwarzchild-anti-de Sitter
  spacetimes. 
  
In simple terms,  these results show that black hole spacetimes formed from the
collapse of topological structures
 must always have both black hole and white hole horizons. This behavior
is quantitatively 
 different  than that of spacetimes containing black holes
formed by collapse of matter which may, but need not exhibit a white hole horizon. 
Therefore, white holes are an essential feature of black hole spacetimes formed from
collapse of topology.

The implications of the corollary for adS/CFT correspondence are immediate. In an
asymptotically de-Sitter spacetime satisfying reasonable physical conditions, any
component of the boundary-at-infinity cannot causally communicate with any other
disjoint component of the boundary-at-infinity.
Thus a field
operator on one component of the boundary-at-infinity cannot causally interact with another field operator
on any other disjoint component. Thus a field operator on one component of
$\scri$ will commute with any other field operator 
on any disjoint component $\scri$.
Thus conformal field theories defined on disjoint components of the boundary-at-infinity do not 
interact dynamically.

Clearly however, one can set up  correlations in the initial vacuum states of the conformal
field theories. In fact, the necessary appearance of white hole horizons may yield
a  natural way to do so. However, any such correlations are not dynamic.

\section{The topology of regions exterior to black hole horizons in $n+1$ dimensions} 

One can obtain further information about regions exterior to black hole horizons if one 
considers the topology of the intersections  of certain spacelike hypersurfaces 
 with the horizons; those for which this intersection is a set of 
closed spacelike $(n-1)$-manifolds (good cuts of the horizons). A
 characterization of these regions can then be given by the
analysis of the topology of these spacelike hypersurfaces, specifically
 Cauchy surfaces as defined for asymptotically anti-de Sitter spacetimes, 
  in terms of their homology.
One can show that in $n+1$ dimensions, 
 spacetimes that obey topological censorship must have spacelike surfaces,
  which contain no wormholes or
 other non-simply connected compact topological structures.
These results are the generalization of the previous results for $3+1$ dimension
reported in \cite{sw:gallowayetal99}  to arbitrary dimension.

Precisely, let ${\cal M'}$ be as described in theorem 1, and $\cal D'$ be the
domain of outer communications of a component  $\scri_0$ of its timelike boundary. 
 A {\it Cauchy surface} (for asymptotically anti-de Sitter spacetime with boundary-at-infinity) 
for $\cal D'$ is defined to 
be a subset $V'\subset \cal D'$ which is met once and only once by each 
inextendible causal curve in $\cal D'$. Then $V'$ will be a spacelike 
hypersurface which, as a manifold-with-boundary, has boundary on $\scri$. 
This definition of Cauchy surface 
for asymptotically anti-de Sitter spacetimes with boundary-at-infinity parallels
that for spacetimes without boundary. For brevity, we  will call these  Cauchy
surfaces for asymptotically anti-de Sitter spacetimes with boundary-at-infinity simply
Cauchy surfaces in the remainder of the paper.

It can be shown, as in the standard case of spacetime without boundary, 
that a spacetime-with-timelike-boundary
$\cal M'$ which is globally hyperbolic admits a Cauchy surface $V'$ for 
$\cal D'$.  Furthermore,  $\cal D'$ is 
homeomorphic to $R\times V'$. (This can be shown by directly modifying the 
proof of Prop.~6.6.8 in \cite{HawkingandEllis}.)  
 Examples of spacetimes whose domains of
outer communications admit such Cauchy surfaces are  
the locally anti-de Sitter spacetimes and related models given in 
\cite{sw:banados1992,sw:lemos95,sw:aminneborgetal96,mann97,brill97}.

If a globally hyperbolic $\cal D'$ has topology
$R\times V'$ (e.g. if $V'$ is a Cauchy surface for $\cal D'$) then
$\cal D'$ can be continuously deformed to $V'$ so that $\scri$ gets
deformed to $\Sigma_0$.  This process preserves fundamental groups
and hence allows the application of
algebraic topology to further characterize the
the spatial hypersurface $V'$. Precisely, 

\noindent {\bf Theorem 4}. {\sl 
Assume $\cal D'$ ($={\cal D}\cup \scri$) is a globally hyperbolic spacetime that
satisfies the conditions given in theorem 1. 
Suppose $V'$ is a Cauchy surface for $\cal D'$ such that its closure 
$V=\overline{V'}$ in $\cal M'$ is a compact topological 
$n$-manifold-with-boundary whose boundary
$\partial V$ (corresponding to the edge of $V'$ in $M'$) consists of
a disjoint union of compact $(n-1)$-manifolds,
$\displaystyle \partial V =\sqcup_{i=0}^{k}\Sigma_i\,$
where $\Sigma_0$ is on  $\scri$ and the $\Sigma_i$,
$i=1,\dots,k$, are on the event horizon.
Then the group homomorphism
$i_*:\Pi_1(\Sigma_0)\to\Pi_1(V)$ induced by inclusion
$i:\Sigma_0\to V$ is onto.}
\vskip 1pt

The proof of this result is given in \cite{sw:gallowayetal99}. Clearly, theorem 4
implies that if $\Sigma_0$ is simply connected, then so is $V$. 

 One may gain further insight into
the consequences of topological censorship by asking how
can one modify the  topology of $V$ yet still satisfy the restriction
$\Pi_1(\Sigma_0)\to\Pi_1(V)$?
 A standard method of constructing topological spaces
 is by connected sum: one takes $V$ and sews in a closed $n$-manifold
$N$  by removing one $n$-ball from the interior of $V$ and one $n$-ball from $N$ then
identifying the resulting $(n-1)$-sphere boundaries with each other to form $ \tilde V =V\# N$. For
 $N=S^{n-1}\times S^1$, this procedure adds a $n$-handle or wormhole to the 
space. One can similarly add another compact connected topological structure to the space by choosing
$N$ to be any other closed manifold besides $S^n$.
 
 Clearly this procedure can add factors that will modify $V$
such that it will no longer satisfy $i_*:\Pi_1(\Sigma_0)\to\Pi_1(V)$ being onto. For example,
the 
addition of a wormhole
will produce a new generator of $\Pi_1(\tilde V)$; a curve passing through
the new handle will not be homotopic to any curve in  $V$. Similarly, adding
another compact connected topological structure which has nontrivial $\Pi_1(N)$ will also introduce new generators 
to $\Pi_1(\tilde V)$. However, connected sums involving compact simply connected topological structures
will not change the fundamental group and thus the addition of such structures is not constrained by
this argument.

An alternate view of this effect is provided in terms of the homology of $V$. For example, taking the connected sum
of $V$ with a handle  $N$ introduces a $(n-1)$-sphere that does not bound an $n$-ball. Therefore,
the rank of $H_{n-1}(V\# N;Z)$ is greater than that of $H_{n-1}(V;Z)$. 
It is  clear on an intuitive level that if connected sums with wormholes and other compact
connected topological structures
can change the topology of $V$, then 
the information provided in theorem 4  must constrain the number of wormholes and other 
compact non-simply connected
 topological structures. 
 
 A further characterization of the spacetime is given by the
analysis of the topology of $V$ in terms of its homology.
 In the following we will
assume that $\Sigma_0$ is orientable, the generalization to the non-orientable case being
straightforward.

\noindent
{\bf Theorem 5}. {\sl If $\Sigma_0$ is orientable and $i_*:\Pi_1(\Sigma_0)\to \Pi_1(V)$ is onto,
then the natural homomorphism $H_1(\Sigma_0;Z)\to H_1(V;Z)$ is onto. 
The integral homology $H_k(V;Z)$ is torsion free for $k=0$, $n-2$, 
$n-1$, and $n$. Furthermore, $H_{n-1}(V;Z)=Z^k$ where $k$ is the 
number of boundaries $\Sigma_i$ interior to $\Sigma_0$.}

\vskip 1pt

\noindent
{\bf Proof}:
We use the fact that the first integral homology group of a space
is isomorphic to the fundamental group modded out by its
commutator subgroup. Hence, modding out
by the commutator subgroups of $\Pi_1(\Sigma_0)$
and $\Pi_1(V)$, respectively, induces from $i_*$ a
surjective homomorphism from $H_1(\Sigma_0;Z)$ to $H_1(V;Z)$.

We next prove the torsion free claims.  The assumption on fundamental groups
and the orientability of $\Sigma_0$ imply that $V$ is orientable.\footnote{ Theorem 2 contradicts
the possibility of a
 nonorientable ${\cal D}'$ satisfying the conditions of theorem 2 with orientable $\scri$. 
The orientable double cover of ${\cal D}'$ would also satisfy the conditions of theorem 2
 and would contain two copies of $\scri$ connected by
causal curves. As this cannot happen,
${\cal D}'$ must be orientable for orientable $\scri$. As orientable $\Sigma_0$ implies
orientable $\scri$, it follows that $V$ is orientable. }
Then, since $V$ has boundary, $H_n(V;Z)=0$.  Also $H_0(V;Z)=Z$ as $V$ is connected. Further,
it is a standard fact that $H_{n-1}(V;Z)$ is free (cf. \cite{Massey}, p. 379).
This follows from Poincar\'e duality for manifolds-with-boundary and the
fact that the relative first cohomology group is in general free.
Thus we need to show that $H_{n-2}(V;Z)$ is free.  The arguments we use
for this we also use to show $H_{n-1}(V;Z)=Z^k$.

\noindent
{\bf Lemma}. {\sl $H_{n-2}(V;Z)$ is free.}

\smallskip
\noindent
To prove the lemma we first consider the relative homology sequence for
the pair $V\supset \Sigma_0$,
\begin{equation}
\cdots\quad\rightarrow H_1(\Sigma_0){\buildrel\alpha\over\rightarrow}
H_1(V){\buildrel\beta\over\rightarrow} H_1(V,\Sigma_0)
{\buildrel\partial\over\rightarrow}{\tilde H_0}(\Sigma_0)=0\quad 
\end{equation}
where we have assumed in the above sequence and from now on in the relative homology arguments
that the
coefficients are over $Z$. Here $\tilde H_0(\Sigma_0)$ is the reduced zeroth-dimensional homology
group. Since, as discussed previously, $\alpha$ is onto,  we have
$\ker \beta ={\rm im}\alpha=H_1(V)$ which implies
$\beta
\equiv 0$. Hence $\ker\partial = {\rm im}\beta =0$, and thus $\partial$
is injective. This implies that $ H_1(V,\Sigma_0)=0$.

Now consider the relative homology sequence for the triple $V\supset
\partial V\supset \Sigma_0$,
\begin{equation}
\cdots\quad\rightarrow H_1(\partial V,\Sigma_0) \rightarrow H_1 (V,\Sigma_0)
=0\rightarrow H_1(V,\partial V) {\buildrel\partial\over\rightarrow}
H_0(\partial V,\Sigma_0)\rightarrow \cdots\quad .\label{7}
\end{equation}
Since $H_0(\partial V,\Sigma_0)$ is torsion free and
$\partial$ is injective, $H_1(V,\partial V)$ is torsion free.
Next, Poincar\'e-Lefschetz duality gives 
$H^{n-1}(V)\cong H_1(V,\partial V)$. Hence $H^{n-1}(V)$ is torsion free.
The universal coefficient theorem implies that
\begin{equation}
H^{n-1}(V)\cong {\rm Hom}\, (H_{n-1}(V),Z)\oplus {\rm Ext}\, (H_{n-2}(V),Z)\quad .
\end{equation}
The functor ${\rm Ext(-,-)}$ is bilinear in the first argument with 
respect to direct sums and ${\rm Ext(Z_k,Z)}=Z_k$. Hence $H^{n-1}(V)$ cannot 
be torsion free unless $H_{n-2}(V)$ is.  This completes the proof of
the lemma.

The boundary surfaces $\Sigma_1$, $\Sigma_2$, $\dots$,
$\Sigma_k$ clearly determine $k$ linearly independent $(n-1)$-cycles in $V$,
and hence $b_{n-1}\ge k$. 

It remains to show that $b_{n-1}\le k$. Since both $H_{n-1}(V)$ and $H^{n-1}(V)$ are
finitely generated and torsion free, we have $H_{n-1}(V)\cong H^{n-1}(V)\cong H_1(V,\partial V)$,
where we have again made use of Poincar\'e-Lefschetz duality.
Hence, $b_{n-1} = {\rm rank}\, H_1(V,\partial V)$.  To show that
${\rm rank}\,H_1(V,\partial V)\le k$, we refer again to the long exact
sequence (\ref{7}).
By excision,
$H_0(\partial V,\Sigma_0) \cong H_0(\partial V \setminus \Sigma_0,\emptyset)
=H_0(\partial V \setminus\Sigma_0)$.  Hence,
by the injectivity of $\partial$,  ${\rm rank}\,H_1(V,\partial V)$ $\le
{\rm rank}\, H_0
(\partial V,\Sigma_0)=$ the number of components of $\partial V \setminus
\Sigma_0 = k$.  This completes the proof of theorem 5. \hfill\endproof
\smallskip

An easy consequence of theorem 5 is that $b_1(\Sigma_0)\geq b_1(V)$.
As observed earlier, the addition of a wormhole changes
not only $\Pi_1(V)$ but also $H_{n-1}(V)$. Also note that rank $H_{n-1}(V;Z) = k$, the
number of boundaries of $V$ interior to the cut of the
boundary-at-infinity $\Sigma_0$. Therefore, there is no element of $H_{n-1}(V;Z)$ associated
with a structure in the interior of $V$. That is, there are no wormholes or other compact
non-simply
connected topological structures in $V$. 

\noindent
{\bf Corollary}. {\sl Given $V$ satisfying the conditions of theorem 5, then there exists
no closed manifold $N$ with $b_1(N)> 0$ such that $V = U\# N$.}

\vskip 1pt
\noindent {\bf Proof}: Observe that as $N$ is a closed manifold, that $U$ has the same boundaries as $V$; 
it follows  that $b_{n-1}(U)\ge k$.  The Mayer-Vietoris sequence yields 
\begin{eqnarray}
0 \rightarrow H_n(V) \rightarrow H_{n-1} (S^{n-1})
&\rightarrow H_{n-1} (U-B^n) \oplus H_{n-1} (N-B^n) \nonumber\cr
&\ \ \ \ \ \ \ \ \rightarrow H_{n-1}(V)\rightarrow
H_{n-2}(S^{n-1} )\cdots\quad .\label{8}
\end{eqnarray}
Now the above sequence is exact as $H_{n-2}(S^{n-1} )=0$. The alternating sum of the ranks must
vanish, thus
$1-(b_{n-1}(U)+1) - b_1(N)) + k = 0$ using $b_n(V)=0$, $b_{n-1}(U-B^n)=b_{n-1}(U)+1$ as $U$ and
$V$ are manifolds with boundary. Clearly this implies $b_{n-1}(U)+b_{n-1}(N)=k$. But this is a
contradiction to $b_1(N)> 0$. \endproof

Finally, one might be worried that  Cauchy surfaces satisfying theorem 4
only occur in stationary spacetimes, that is
where no black hole formation occurs. However, this is not the case as first pointed out
by Jacobson and Venkataramani \cite{sw:jacobsonetal95}; one can construct surfaces that characterize
cuts of black hole horizons that occur via collapse in quite general asymptotically flat
 spacetimes. What one does is
construct a globally hyperbolic spacetime  that consists of
a subset of the original spacetime. One can carry out a similar construction
for  the asymptotically anti-de Sitter spacetimes. 
 Precisely, let
 $K$ be a cut of $\scri$, and let $\scri_K$ be
the portion of $\scri$
to the future of $K$, $\scri_K = \scri \cap I^+(K)$.  Let  ${\cal D}_K$ be
the domain of
outer of communications with respect to $\scri_K$,  ${\cal D}_K =
I^+(\scri_K)\cap I^-(\scri_K)
= I^+(K) \cap I^-(\scri)$. One chooses ${\cal D}_K'={\cal D}_K\cup\scri$ such that 
it is globally hyperbolic and such that the closure of its Cauchy surface in ${\cal M'}$
has a good intersection with the black hole horizons, i.e. the intersections 
are $(n-1)$-manifolds. This new spacetime satisfies the conditions of theorem 1 and
will contain a surface $V$ as required in theorem 4. Therefore the conclusions of 
theorem 5 hold for such surfaces as well.
Thus, though topological censorship does not determine the topology of arbitrary
embedded hypersurfaces, it does
do so for hypersurfaces 
homeomorphic to Cauchy surfaces for the domain of outer communications that make good cuts
of the horizons. Details regarding this procedure
are discussed further 
in \cite{sw:gallowayetal99}.

\section{Further results in $2+1$  and $4+1$  dimensions }

Clearly, by using special
properties of manifolds in a given dimension $n$, the results obtained here may
be strengthened. This is particularly true in low dimension. Of special relevance to the 
adS/CFT correspondence conjecture
are results on asymptotically anti-de Sitter spacetimes in $2+1$ and $4+1$ dimensions.

In three dimensions one can show that

\noindent
{\bf Theorem 6}. {\sl  Assume $\cal D'$ is a globally hyperbolic spacetime-with-boundary that
satisfies the conditions of theorem 4. Then the 2-dimensional hypersurface $V$ 
is either $B^2$ or $I\times S^1$.}

\vskip 1pt
\noindent
{\bf Proof}: As all 1-manifolds are orientable, $V$ is orientable. 
Theorem 5 implies that the rank of the free part of $H_1(V;Z)$
cannot be greater than that of $H_1(\Sigma_0;Z)$, {\it i.e.},
$b_1(V) \le b_1(\Sigma_0)$ In the case $n=2$, $\Sigma_0$ is
a 1-manifold so  $b_1(\Sigma_0)\le 1$ thus $b_1(V)\le 1$.  Now $V$ is a closed 2-manifold minus a
disjoint union of discs.
From the classification of 2-manifolds, $V$ must be a closed 2-manifold  minus a disjoint
union of disks. The first betti number of such manifolds is
$b_1=2g+k$ where $g$ is the genus and $k+1$ the number of disjoint disks; it follows that $g=0$.
 Since $V$ must have 
at least one boundary,   the only 
possible topologies for $V$ are $B^2$ or $I\times S^1$.
\hfill\endproof

Theorem 6 has very interesting consequences for the topology of $(2+1)$-dimensional 
spacetimes.  If $\scri$ is disconnected, then $V'$ for
the domain of outer communications of each disconnected component of $\scri$ 
will have product topology. Thus topological censorship gives a topological
rigidity theorem in $(2+1)$-dimensional gravity. 
As one has directly characterized the topology of the domain of outer communications
for these spacetimes, it follows,
by arguments  similar to that used to characterize the topology of good cuts of black 
hole horizons in the $(3+1)$-dimensional case given in \cite{sw:gallowayetal99}, 
that the topology of a good cut of a black hole
horizon in $(2+1)$-dimensional spacetime is always $S^1$. 

The case of $(2+1)$-dimensional asymptotically flat spacetimes
 can be similarly treated to produce the
same conclusions as theorem 6. It follows that
there are no asymptotically flat geons in
three dimensions.

In the case of $(4+1)$-dimensional spacetimes, theorem 5 yields that the
integral homology $H_k(V;Z)$ is torsion free except for $k=1$. However, theorem 5 and the
onto homomorphism of the fundamental groups is not enough to even partially fix the topology of
$V$. To demonstrate this, it is useful to first study the restricted case for which
$\Sigma_0$ is simply connected. It follows that $V$ is 
 a simply connected
manifold with boundary. This is a fairly significant restriction; however one will have an
infinite number of such manifolds. One obtains these simply by
taking the connected sum of $V$ with any closed simply connected 4-manifold. One can 
readily show that
the connected sum of two such manifolds leaves $H_k$ unchanged except for $H_2$. There are 
an infinite number
of closed simply connected $4$-manifolds characterized by their Hirzebruch signature 
and Euler characteristic. 

Furthermore the restriction that $V$ is simply connected is not enough to deduce the topology of
the boundaries $\Sigma_i$ even in this simple case. It is well 
known that all closed 3-manifolds
are cobordant to $S^3$.  In fact 
one can construct a cobordism with trivial fundamental group \cite{trivial}.  Therefore, one
cannot conclude any restriction on the topology of the cuts of  black hole horizons in
$(4+1)$-dimensional spacetimes from the 
simple arguments given above.

 Finally, it is clear  that similar conclusions follow in the case of
  non-simply connected $\Sigma_0$. In particular,  one will have an
infinite number of  manifolds with the same fundamental group and $H_{n-1}(V;Z)$ obtained
 by
taking the connected sum of $V$ with any closed simply connected 4-manifold. Thus the topology
of the interior of a $(4+1)$-dimensional asymptotically de Sitter spacetime is constrained but
not completely characterized by the topology of the boundary-at-infinity.

\acknowledgements

We would like to thank G. Semenoff and E. Witten for useful conversations. 
This work was partially supported by
the Natural Sciences and Engineering Research Council of Canada and
by the National Science Foundation (USA), Grant No.~DMS-9803566.

\end{document}